\documentstyle[preprint,prc,aps]{revtex}

\begin{document}
\draft

\title{$\pi^+ + d \rightarrow p+p$ reaction between 18 and 44~MeV}
\author{E.~A.~Pasyuk, V.~Yu.~Alexakhin, S.~I.~Gogolev, K.~O.~Oganesyan}
\address{Joint Institute for Nuclear Research, Dubna, Moscow region,
141980 Russia}
\author{ C.~L.~Morris, J.~M.~O'Donnell\cite{JMOD}, M.~W.~Rawool-Sullivan}
\address{Los Alamos National Laboratory, Los Alamos, New Mexico 82740}
\author {M.~K.~Jones\cite{MKJ}}
\address{Rutgers Univesity, Piscataway, New Jersey 08855}
\author{F.~F.~Guber, A.~I.~Reshetin}
\address{Institute for Nuclear Research, Troitsk, Moscow region, 117312 Russia}
\author{I.~I.~Strakovsky}
\address{St.~Petersburg Nuclear Physics Institute, Gatchina,
St.~Petersburg, 188350 Russia}

\date{\today}

\maketitle

\begin{abstract}

A study of the reaction $\pi^+ + d \rightarrow p+p$ has been performed
in the energy range of 18 -- 44~MeV. Total cross sections and
differential cross sections at six angles have been measured at 15
energies with an energy increment of 1 -- 2~MeV.  This is the most
systematic data set in this energy range. No structure in the energy
dependence of the cross section has been observed within the accuracy
of this experiment.

\end{abstract}

\pacs{PACS number(s): 25.80.Ls, 13.75.Cs 13.75.Gx, 14.20.Pt}

\narrowtext

\section{Introduction}

Total and differential cross sections for the reaction $\pi^+ d \to
pp$ have been measured with high accuracy at pion energies above
100~MeV.  Currently, experimental efforts in this energy range are
aimed at measuring spin dependent observables.  The situation at pion
energies below 100~MeV is less complete, especially for energies of a
few tens of MeV\cite{Weyer}. The disagreement between the experimental
data for the total cross section in some cases exceeds 20\%, much
larger than the quoted uncertainties. The experimental data show a
deviation from smooth behavior in the total cross section of the
reaction $\pi^+ d \to pp$ close to $T_{\pi}\sim 30\ \text{MeV}$
($\sqrt{s} = 2.04\ \text{GeV/c}^2$)\cite{ISEP}. More recent
measurements of the partial total cross section for the inverse
reaction at SATURNE have verified this effect\cite{SATURN}.  In a
recent experiment of the Dubna-Gatchina group on pion absorption on
carbon\cite{Akimov} a dip in the energy dependence of the
quasi-deuteron component of absorption near $T_{\pi}$=28~MeV has been
observed.  The measurements of pion absorption on the
deuteron\cite{Gogolev} performed by the same group also indicated
possible structure at a pion energy of 30~MeV.  A more recent
experiment at LAMPF\cite{Jones} did not observe any dip on carbon, but
these data suggest a change in the angular distribution of protons at
pion energies near 25~MeV.  Some possible alternative explanations of
the structure in this energy range are given in
Ref.\cite{ISEP,Akimov}.  One of them attributed the structure to the
excitation of a diproton resonance in the $^3P_2 \quad NN$ state.
However, these experiments are not accurate enough to reach a
conclusion concerning the structure under the discussion.  A
satisfactory solution fitting the available database has not been
found in partial wave analysis, suggesting underlying systematic
problems with the data base.

The experiment described here was intended to resolve the
discrepancies in existing data and to verify the observed structure in
the excitation function of pion absorption on the deuteron at low pion
energies.  The differential and total cross sections of the $\pi^+ d
\to pp$ reaction have been measured with fine steps in incident pion
energy.

\section{Experiment}

\subsection{Pion Beam and Target}

The experiment was performed at the Low Energy Pion channel (LEP) of
the Clinton P.~Anderson Meson Physics Facility (LAMPF).  Positive
pions with energies of 21, 23, 25, 26, 27, 28, 29, 30, 31, 33, 35, 37,
39, 41 and 45~MeV were used.  The pion beam had an average intensity
of a few 10$^4$/sec, allowing the beam particles to be counted while
not overloading the BGO detectors. The pion fraction varied from 78\%
for 45~MeV beam to 33\% for 21~MeV beam. The momentum bite of the LEP
channel was set to 1\% for the highest energies and increased to 4\%
for the lowest ones in order to increase the pion flux.

The target was composed of CD$_2$ with a cross sectional area of
$1\times 1 \text{cm}^2$ and an areal density of 0.469~g/cm$^2$.  It
was attached to a thin paper pipe and placed in the center of the BGO
ball.  The supporting pipe was aligned along the beam axis. The
diameter of the pipe was big enough to keep its walls out of the beam.
A 0.25-mm-thick plastic scintillator S1 with a cross section of
$6\times 6 \ \text{mm}^2$ was located just before the target.
Downstream of the target a rectangular array of nine CsI scintillators
DA1--DA9 was preceded by a 10-mm-thick plastic scintillator, S2.
DA1--DA9 and S2 could be used for the detection of the most forward
going reaction products as well as for a determination of beam
composition. In this experiment only the central detector, DA5, of the
array was used.  A coincidence between S1 and the central detector,
DA5, of the downstream array was used as a beam monitor. Every 1000'th
beam event (event triggered by a $S1 \cdot DA5$ coincidence) was read out
for further analysis to determine the pion fraction which was needed
for absolute normalization of the cross sections.

\subsection{BGO Ball Spectrometer}

A large solid angle detector, the LAMPF BGO ball, was used to detect
the reaction products in this study. Detailed information on the BGO
ball can be found in Ref. \cite{{bgo1},{bgo2}}. The BGO ball consists
of 30 phoswich detectors.  The detectors of the array were of
pentagonal and hexagonal shape and tightly packed to form a truncated
icosahedron of 32 sides. Two of the 32 sides are opened for the beam
entry and exit. The detectors were distributed about an inner radius
of 6.1~cm from the center of the array to the center of each crystal
face, and were arranged in six groups centered at laboratory
scattering angles of $\theta = 37^{\circ}, 63^{\circ},79^{\circ},
102^{\circ}, 116^{\circ}$, and $142^{\circ}$. Each detector had a
solid angle of about $\frac {1}{32} \times 4\pi$ sr and was supported
in a 0.5-mm-thick electro-formed nickel can which had a 0.05-mm-thick
entrance window.  Each detector consisted a 3-mm-thick NE102 plastic
scintillator optically coupled to the front of a 5.6-cm-thick bismuth
germanate (BGO) crystal, with a 7.62-cm-diameter photomultiplier tube
on the back. Since the decay constant of the BGO scintillator is much
longer than that of the plastic scintillator (250~ns vs 1.5~ns), the
anode signal was time sliced to provide both $\Delta E$ (fast) and $E$
(slow) signals for charged particle identification (pions, protons,
deuterons, etc.), and for identification of neutrons and gamma
rays. The crystals were thick enough to stop up to 185-MeV protons and
90-MeV pions. The time resolution of the detectors was about 1~ns,
sufficient to eliminate events with hits from different beam bursts
(the LAMPF beam has a 5-ns microstructure). The light output of BGO
scintillator depends significantly on the temperature of BGO
material\cite{bgo3}. To minimize fluctuations in temperature of the
BGO, a tent-like structure was built to isolate the BGO ball from its
surroundings.

The event trigger consisted of a coincidence between the target detector,
S1, and at least one BGO crystal in anti-coincidence with  DA5.

\section{Data Analysis Procedure}

\subsection{Raw Data Processing}

The raw data for each event contains information about the energy
deposited in the plastic and the BGO for all detectors and timing
information with respect to the beam counter. In the first step of the
analysis a time gate was applied to all of the signals to remove
accidentals from further analysis.  The next step was the
determination and application of the constants used for unmixing the
${\Delta}E - E$ information from the phoswich detectors. The signal
from each phoswich detector was integrated in two different ADC
channels with different time gates, 50 and 250~ns long.  The $\Delta
E$ and $E$ information was separated using:

\begin{equation} \FL \Delta E_i =(dE_i - E1_i \cdot R_{i,1} - Z_{i,1})
\cdot DEGAIN_i \quad , 
\end{equation} 
\begin{equation} E_i = (E1_i - dE_i \cdot
R_{i,2} - Z_{i,2}) \cdot EGAIN_i \quad , 
\end{equation} 

\noindent where $dE_i$ and $E1_i$ represent the raw data from the ADC's with the
short and long gates, respectively, and $DEGAIN_i$, $EGAIN_i$ are the
coefficients for conversion of raw ADC data to energy in MeV. The
relative fraction of a long signal in a short gate and vice versa are
given by the mixing parameters, $R_{i,1}$ and $R_{i,2}$,
respectively. The $R_{i,j}$ and offsets, $Z_{i,j}$, were determined
using a least square fit for the events along $\Delta E$ and $E$ axes,
where only one component of a signal (short or long) exists. These are
produced by charge particles which stop in the plastic and leave a
$\Delta E$ with no $E$ signal, or by neutral particles which interact
in the BGO and leave a $E$ with no $\Delta E$ signal.  An example of a
${\Delta}E-E$ distribution from a phoswich is shown in
Fig. \ref{ede}. One can clearly see regions corresponding to protons
and pions. The outlined region along the $E$ axis is due to neutral
particles and the outlined region along the $\Delta E$ axis is due to
short range charged particles stopped in the plastic scintillator.

\subsection{Calibration and Stabilization of the BGO Ball}

An initial energy calibration of the BGO ball ($DEGAIN_i$ and
$EGAIN_i$) was made by using two proton coincidences from the $\pi^+ d
\rightarrow pp$ reaction. An alternate method used elastically
scattered pions from $^{12}C$.  Pions from this reaction give a strong
peak (Fig. \ref{ede}) with well defined energy for each ring of
detectors in the BGO ball. This eliminated the strong kinematic energy
dependence in the former procedure.  Use of the CD$_2$ as a target and
a single hit trigger made this possible. The $\Delta E$ and $E$ gains
were fitted continuously during data taking and off-line data
analysis.  This continuous stabilization was important because of
instability in the BGO ball detectors.  The parameters $R_{i,j}$ and
$Z_{i,j}$ also require continuous stabilization because both the gain
and the decay constant of the BGO scintillator is temperature
dependent. We used the following stabilization algorithm.  For
sequential subsets of raw events the complete set of parameters was
fitted. The weighted average of old and new values of the parameters
was used for the next subset of events. This procedure was completely
automated. Fig. \ref{pp} (described below) shows the energy resolution
and accuracy of this calibration procedure. The width of the two
proton total energy peak from the $d(\pi^+,p)p$ reaction, summed over
entire ball is $\sigma =3 - 4\%$ MeV. After the system has been
calibrated one can easily determine the event multiplicity, identify
detected particles and measure their energies.

\section{results}

\subsection{Event Selection Criteria}

We selected events using following criteria:
\begin{itemize}
\item[(1)]
the multiplicity must be equal 2;
\item[(2)]
both particles must be protons; and
\item[(3)]
the opening angle between the two protons and their total energy
must satisfy the kinematics of the reaction $\pi^+ d \rightarrow pp$.
\end{itemize}

\noindent
Fig. \ref{pp} is the energy spectrum for two proton events. The events
are summed over all kinematically allowed combinations of BGO ball
detectors. The narrow peak at the higher energy corresponds to the
reaction $\pi^+ d \rightarrow pp$. The lower energy broad peak with a
long tail results from pion absorption on carbon. One observes that
there is practically no background under the $\pi^+ d \rightarrow pp$
reaction peak. Measurements with no target but with the supporting
pipe in its normal position and with a pure carbon target show that
the background under the peak does not exceed 0.1\% so no background
subtraction was necessary.

\subsection{Pion Fraction Determination}

Because of the low beam flux in this experiment it was possible to
directly count the beam particles.  Pulse height and timing
information from S1, S2 and DA5 detectors was used to determine the
pion fraction.  First, a cut was placed on the time between S1 and S2
to eliminate accidental coincidences between particles from different
beam bursts.  Another cut was applied to a two dimensional $\Delta E -
E$ distribution (pulse height from S2 and DA5) to eliminate
positrons. The resulting pulse height spectrum from S2 is shown in
Fig.  \ref{beam}. The left peak corresponds to muons, the right one to
pions.  Two modified Moyal functions:\cite{Moyal}

\begin{equation}
F(E)=P_{1} e^{(-P_{3}(E-P_{2}) - e^{-P_{4}(E-P_{2})})} , \label{eMoyal}
\end{equation}

\noindent one for pions and one for muons, were fitted to the energy
loss distribution.  Here $E$ is the energy loss in S2 and $P_{1} -
P_{4}$ are adjusted parameters.  The pion distribution was integrated
to obtain the pion fraction.  At the two lowest energies (21~MeV and
23~MeV) some pions loose all their energy in S2 and do not hit DA5 and
the corresponding pion peak has a different, and more complicated
shape. In this case only the muon peak was fitted and the pion
fraction was determined by subtracting the muon fraction from the
integral number of beam particles.  A Monte Carlo simulation of beam
particles passing through the setup based on the GEANT
code\cite{GEANT} has been done.  The corrections for energy losses,
straggling, multiple scattering and decay were calculated from this
simulation. The resulting correction to the total number of beam pions
is 7\% for the highest beam energy and 44\% for the lowest one.  For
beam energies of 21 and 23~MeV this correction accounts for pions,
which passed through the target but did not hit DA5 and were lost from
the beam trigger. This simulation also was used for calculation of the
pion energy in the center of a target.  The uncertainty in the number
of beam pions, mainly due to fitting errors, is 3 -- 5\%, with the
contribution from the statistical error less than 1\%.

\subsection{Efficiency Calculation}

The efficiency was calculated using a Monte Carlo code which
incorporates the actual geometry of the experimental setup.  The
efficiency accounts for the effective solid angle (geometrical solid
angle reduced by the requirement of both protons detection), reaction
losses, and dead time.  The effective solid angles for all
kinematically allowed pairs of the detectors have been calculated.
The missing solid angle due to the nickel cans surrounding the
crystals is about 6\%. Reaction loss correction due to the interaction
of protons with the BGO were obtained using data from
Ref. \cite{bgo4}.  The typical value of this correction is 10\%.
Finally, the data were corrected for dead time.  A typical value of
the dead time in this experiment was 5\%.  We estimate the total
uncertainty in the efficiency of the setup is about 5\%.

\subsection{Cross Sections}

The differential cross section has been calculated for the selected
events at each specific scattering angle. The data for the backward
angles were translated to the forward hemisphere since the angular
distribution in the center of mass frame is symmetrical around
$90^{\circ}$ due to indistinguishability of the two protons. The
resulting differential cross sections are presented in Table
\ref{tdsig}. To obtain the total error, we added the statistical and
normalization errors in quadrature. An overall systematic uncertainty
of about 5\% in the efficiency calculation is not included in the
total error. This affects the global normalization of all 90
differential cross section points and does not change their relative
value.

We fitted the cross sections using a simple parameterization given by
the Legendre polynomial series:

\begin{equation} 
\frac{d\sigma}{d\Omega_{c.m.}} = a_{0} + a_{2}P_{2}(\cos
{\theta}_{c.m.}). \label{Leg} 
\end{equation} 

\noindent At each energy we used only the statistical errors of the
cross sections but not the normalization error because it does not
change the relative shape of the angular distribution.

The total cross sections were determined by:

\begin{equation}
\sigma_{tot} = \int_{2\pi} \frac{d\sigma}{d\Omega_{c.m.}} d\Omega_{c.m.}
\end{equation}

\noindent using the parameterization of Eq. (\ref{Leg}). The
integration over $2\pi$ instead of $4\pi$ is necessary because there
are two indistinguishable protons in the final state.  The fitted
coefficients, total cross sections, and $\chi ^{2}$ are presented in
Table \ref{tsgt}.  Errors for $a_0$ and $a_2$ are obtained from the
fit. For the total cross section errors we added the normalization
errors in quadrature.  Overall systematic uncertainty of 5\% is not
included for the same reason as for differential cross section.

\section{Discussion}

The differential cross sections measured in this experiment are
plotted in Fig.~\ref{dsig}. Solid lines show the best fit results with
Eq.~(\ref{Leg}).  The results listed in Table~\ref{tsgt} show that
only S- and P- pionic waves are important in this energy region.  The
absence of a need for higher partial waves agrees with previous
measurements\cite{Ritchie,Mathie}. Fixed angle excitation functions
are presented in Fig.~\ref{dsigr}. Fig.~\ref{a2a0} shows the behavior
of the angular distribution as a function of energy by plotting the
ratio of the coefficients $a_2$ to $a_0$. The shape of the angular
distribution changes smoothly with pion energy.  No structure, within
the accuracy of the experiment, is observed in the differential cross
section as a function of energy.  The total cross section is presented
in Fig.~\ref{sgt} along with the rest of the world's data. The current
data follows the general trends.  The biggest disagreement in the
total cross section data, which existed around $T_{\pi}=30$~MeV is
resolved by the new data.

We compared our cross sections with the predictions of the recent
partial wave analysis of Ref. \cite{SAID}. The new VPI solution, SM95,
that includes the current data is compared with the previous SP94
results which do not include these data.

Table~\ref{tab3} and Figs.~\ref{dsigr},\ref{sgt} show the level of
agreement between or data and the phase shift fits.  The normalization
factor averaged over all 15 energies are 1.04 and 1.01 for SP94 and
SM95 solutions respectively.  The values of the normalization factors
for both solutions show that our estimate of systematic uncertainties
is reasonable.  Table~\ref{tab4} presents a comparison of our data
with other partial wave analysis. The recent VPI results, which
include our data set, give better agreement than the previous results
by the Queen Mary College group (9~--~256~MeV)~\cite{bugg} and by the
Hiroshima University group ( 6~--~256~MeV)~\cite{hiro}.  The good
agreement between the predictions of SP94 and recent SM95 VPI
solutions with the experimental data support the absence of any
anomalous behavior of the cross sections in this energy region within
the accuracy of the present experimental results.

\section{Conclusions}

We have presented new data on the total and differential cross section
for the reaction $\pi^+ d \to pp$ for pion energies from 18~MeV to
44~MeV.  In this experiment, 90 experimental data points for the 
differential cross section and 15 points for the total cross section have
been measured with the same systematic uncertainty.  The number of the
experimental data points in this energy range is almost doubled by
these results.  The data generally follow the trends of previous
results and are in a good agreement with the recent SM95 solution of the VPI
partial wave analysis. The previous disagreement in total cross
section data is resolved by this experiment. We do not observe any
structure either in total cross section or in differential cross
section.

\acknowledgments

The authors wish to acknowledge valuable discussions with R.~A.~Arndt,
R.~A.~Giannelli, R.~D.~Ransome, and B.~G.~Ritchie.  This work was
supported in part by Russian Fundamental Research Foundation Grant
93-02-3995 and by the U.S.~Department of Energy.

\begin{figure}

\caption{Contour plot of the two dimensional distribution of $\Delta
E$ (in plastic) vs. $E$ (in BGO) obtained from BGO ball phoswich
detectors.  Regions corresponding to different particle types are
labeled on the plot.}
\label{ede}

\end{figure}

\begin{figure}

\caption{Observed total energy spectrum for two protons produced in
pion absorption on a CD$_2$ target at $T_{\pi}=43.5$ MeV.}
\label{pp}

\end{figure}

\begin{figure}

\caption{Pulse height spectrum from the S2 detector for a 30 MeV pion
beam.  The histogram is a measured spectrum, dashed and dotted lines
represent the fit results with Eq. (\protect\ref{eMoyal}) for muons
and pions respectively. Solid line is a sum of these two
functions. Positrons have been rejected.}
\label{beam}

\end{figure}

\begin{figure}

\caption{Differential cross sections determined in this experiment.
Solid lines are the best fits with Eq. (\protect\ref{Leg}).}
\label{dsig}
\end{figure}

\begin{figure}

\caption{Energy dependence of the differential cross section at fixed
scattering angles. Solid lines are the predictions from the recent VPI
solution SM95 \protect\cite{SAID}. }
\label{dsigr}
\end{figure}

\begin{figure}
\caption{Ratio of the coefficients of a Legendre polynomial expansion
of the differential cross section with Eq. (\protect\ref{Leg}).}
\label{a2a0}
\end{figure}

\begin{figure}

\caption{Total cross section. Black circles are the data from this
experiment. Open circles represent previous experimental data set for
this energy range (taken from SAID \protect\cite{SAID} data
base. Solid line is the VPI SM95 global fit to data in the range 0 --
550~MeV \protect\cite{SAID}.}
\label{sgt}
\end{figure}

\newpage
\begin{table}
\caption{Differential cross sections. Overall systematic uncertainty
of 5\% is not included in the total error.}
\label{tdsig}

\begin{tabular}{cccccc}
$T_{\pi}$\tablenote{Energy in the center of target}&
$\theta_{c.m.}$ & $d\sigma/d\Omega_{c.m.}$ & statistical&
normalization\tablenote{Same value for all six angles at given energy}&
total\tablenote{Total error is the statistical and normalization error
summed in quadrature.} \\
(MeV)& (deg)& (mb/sr)& error& error (\%)&
error \\
\tableline
18.8 & $36^{+\ 7}_{-\ 9}$ & 0.891  & 0.045  & 2.9  & 0.052\\
     & $41^{+\ 9}_{ -12}$ & 0.793  & 0.035  &      & 0.042\\
     & $58^{+\ 9}_{-\ 9}$ & 0.669  & 0.038  &      & 0.042\\
     & $66^{ +10}_{ -10}$ & 0.600  & 0.033  &      & 0.038\\
     & $73^{ +10}_{ -10}$ & 0.460  & 0.030  &      & 0.033\\
     & $83^{ +11}_{ -11}$ & 0.430  & 0.028  &      & 0.031\\
20.9 & $36^{+\ 7}_{-\ 9}$ & 0.860  & 0.043  & 4.2  & 0.056\\
     & $41^{+\ 9}_{ -12}$ & 0.797  & 0.034  &      & 0.048\\
     & $58^{+\ 9}_{-\ 9}$ & 0.638  & 0.036  &      & 0.045\\
     & $66^{ +10}_{ -10}$ & 0.533  & 0.030  &      & 0.038\\
     & $73^{ +10}_{ -10}$ & 0.437  & 0.028  &      & 0.034\\
     & $83^{ +11}_{ -11}$ & 0.365  & 0.025  &      & 0.029\\
22.9 & $36^{+\ 7}_{-\ 9}$ & 0.795  & 0.040  & 4.5  & 0.054\\
     & $41^{+\ 9}_{ -12}$ & 0.710  & 0.031  &      & 0.045\\
     & $57^{+\ 9}_{-\ 9}$ & 0.592  & 0.033  &      & 0.043\\
     & $66^{ +10}_{ -10}$ & 0.490  & 0.028  &      & 0.036\\
     & $73^{ +10}_{ -10}$ & 0.456  & 0.028  &      & 0.035\\
     & $84^{ +11}_{ -11}$ & 0.422  & 0.026  &      & 0.032\\
24.0 & $36^{+\ 7}_{-\ 9}$ & 0.906  & 0.049  & 4.3  & 0.062\\
     & $41^{+\ 9}_{ -12}$ & 0.800  & 0.038  &      & 0.051\\
     & $57^{+\ 9}_{-\ 9}$ & 0.676  & 0.040  &      & 0.050\\
     & $67^{ +10}_{ -10}$ & 0.621  & 0.036  &      & 0.044\\
     & $73^{ +10}_{ -10}$ & 0.458  & 0.032  &      & 0.037\\
     & $84^{ +11}_{ -11}$ & 0.418  & 0.029  &      & 0.034\\
25.0 & $36^{+\ 7}_{-\ 9}$ & 0.874  & 0.045  & 4.1  & 0.057\\
     & $41^{+\ 9}_{ -12}$ & 0.808  & 0.035  &      & 0.048\\
     & $58^{+\ 9}_{-\ 9}$ & 0.674  & 0.037  &      & 0.047\\
     & $68^{ +10}_{ -10}$ & 0.557  & 0.031  &      & 0.039\\
     & $72^{ +10}_{ -10}$ & 0.443  & 0.029  &      & 0.034\\
     & $84^{ +11}_{ -11}$ & 0.374  & 0.026  &      & 0.030\\
26.1 & $36^{+\ 7}_{-\ 9}$ & 0.984  & 0.048  & 3.6  & 0.060\\
     & $41^{+\ 9}_{ -12}$ & 0.897  & 0.037  &      & 0.049\\
     & $57^{+\ 9}_{-\ 9}$ & 0.728  & 0.039  &      & 0.047\\
     & $67^{ +10}_{ -10}$ & 0.615  & 0.033  &      & 0.040\\
     & $72^{ +10}_{ -10}$ & 0.483  & 0.031  &      & 0.035\\
     & $84^{ +11}_{ -11}$ & 0.414  & 0.027  &      & 0.031\\
27.1 & $36^{+\ 7}_{-\ 9}$ & 0.940  & 0.046  & 3.7  & 0.058\\
     & $41^{+\ 9}_{ -12}$ & 0.901  & 0.036  &      & 0.050\\
     & $57^{+\ 9}_{-\ 9}$ & 0.757  & 0.039  &      & 0.048\\
     & $67^{ +10}_{ -10}$ & 0.607  & 0.032  &      & 0.039\\
     & $72^{ +10}_{ -10}$ & 0.530  & 0.031  &      & 0.037\\
     & $84^{ +11}_{ -11}$ & 0.476  & 0.029  &      & 0.034\\
28.1 & $36^{+\ 7}_{-\ 9}$ & 0.948  & 0.049  & 3.7  & 0.060\\
     & $41^{+\ 9}_{ -12}$ & 0.931  & 0.039  &      & 0.052\\
     & $57^{+\ 9}_{-\ 9}$ & 0.767  & 0.042  &      & 0.051\\
     & $67^{ +10}_{ -10}$ & 0.632  & 0.035  &      & 0.042\\
     & $72^{ +10}_{ -10}$ & 0.504  & 0.032  &      & 0.037\\
     & $84^{ +11}_{ -11}$ & 0.421  & 0.029  &      & 0.033\\
29.2 & $36^{+\ 7}_{-\ 9}$ & 0.972  & 0.052  & 3.6  & 0.062\\
     & $41^{+\ 9}_{ -12}$ & 0.965  & 0.041  &      & 0.054\\
     & $57^{+\ 9}_{-\ 9}$ & 0.755  & 0.043  &      & 0.051\\
     & $67^{ +10}_{ -10}$ & 0.557  & 0.034  &      & 0.039\\
     & $72^{ +10}_{ -10}$ & 0.508  & 0.034  &      & 0.038\\
     & $84^{ +11}_{ -11}$ & 0.512  & 0.033  &      & 0.038\\
31.2 & $36^{+\ 7}_{-\ 9}$ & 1.032  & 0.053  & 3.5  & 0.064\\
     & $42^{+\ 9}_{ -12}$ & 0.997  & 0.042  &      & 0.054\\
     & $57^{+\ 9}_{-\ 9}$ & 0.815  & 0.045  &      & 0.053\\
     & $67^{ +10}_{ -10}$ & 0.608  & 0.035  &      & 0.041\\
     & $72^{ +10}_{ -10}$ & 0.490  & 0.033  &      & 0.037\\
     & $85^{ +11}_{ -11}$ & 0.415  & 0.029  &      & 0.033\\
33.3 & $36^{+\ 7}_{-\ 9}$ & 1.043  & 0.043  & 4.2  & 0.053\\
     & $42^{+\ 9}_{ -12}$ & 0.921  & 0.032  &      & 0.042\\
     & $56^{+\ 9}_{-\ 9}$ & 0.703  & 0.033  &      & 0.039\\
     & $67^{ +10}_{ -10}$ & 0.539  & 0.027  &      & 0.031\\
     & $72^{ +10}_{ -10}$ & 0.455  & 0.025  &      & 0.029\\
     & $85^{ +11}_{ -11}$ & 0.409  & 0.023  &      & 0.026\\
35.4 & $36^{+\ 7}_{-\ 9}$ & 1.032  & 0.043  & 4.0  & 0.060\\
     & $42^{+\ 9}_{ -12}$ & 0.929  & 0.032  &      & 0.049\\
     & $56^{+\ 9}_{-\ 9}$ & 0.768  & 0.035  &      & 0.047\\
     & $68^{ +10}_{ -10}$ & 0.611  & 0.028  &      & 0.037\\
     & $71^{ +10}_{ -10}$ & 0.470  & 0.026  &      & 0.032\\
     & $85^{ +11}_{ -11}$ & 0.435  & 0.024  &      & 0.030\\
37.4 & $36^{+\ 7}_{-\ 9}$ & 1.101  & 0.068  & 4.9  & 0.087\\
     & $42^{+\ 9}_{ -12}$ & 0.930  & 0.048  &      & 0.067\\
     & $56^{+\ 9}_{-\ 9}$ & 0.711  & 0.051  &      & 0.062\\
     & $68^{ +10}_{ -10}$ & 0.534  & 0.040  &      & 0.048\\
     & $71^{ +10}_{ -10}$ & 0.447  & 0.038  &      & 0.044\\
     & $85^{ +11}_{ -11}$ & 0.408  & 0.035  &      & 0.041\\
39.5 & $36^{+\ 7}_{-\ 9}$ & 1.205  & 0.055  & 3.4  & 0.068\\
     & $42^{+\ 9}_{ -12}$ & 1.171  & 0.041  &      & 0.057\\
     & $56^{+\ 9}_{-\ 9}$ & 0.961  & 0.045  &      & 0.056\\
     & $68^{ +10}_{ -10}$ & 0.691  & 0.035  &      & 0.042\\
     & $71^{ +10}_{ -10}$ & 0.567  & 0.033  &      & 0.038\\
     & $86^{ +11}_{ -11}$ & 0.456  & 0.028  &      & 0.032\\
43.5 & $36^{+\ 7}_{-\ 9}$ & 1.353  & 0.048  & 3.8  & 0.070\\
     & $42^{+\ 9}_{ -12}$ & 1.201  & 0.035  &      & 0.057\\
     & $56^{+\ 9}_{-\ 9}$ & 0.946  & 0.037  &      & 0.052\\
     & $68^{ +10}_{ -10}$ & 0.732  & 0.029  &      & 0.040\\
     & $71^{ +10}_{ -10}$ & 0.606  & 0.028  &      & 0.036\\
     & $86^{ +11}_{ -11}$ & 0.515  & 0.025  &      & 0.032\\
\end{tabular}
\end{table}

\newpage
\begin{table}

\caption{The Legendre polynomial expansion coefficients of the
differential cross section as fit with Eq. (\protect\ref{Leg}) and the
associated total cross sections. Errors for $a_0$ and $a_2$ are
obtained from the fit.  For the total cross section error the
normalization errors are added in quadrature.  The overall systematic
uncertainty of 5\% is not included.} \label{tsgt}

\begin{tabular}{ccccc}
$T_{\pi}$  & $a_0$   & $a_2$   & $\chi^2/data$ & $\sigma_{tot}$ \\
(MeV)      & (mb/sr) & (mb/sr) &               &     (mb)       \\
\tableline
18.8 & 0.663$\pm 0.015$ & 0.458$\pm 0.041$ & 4.7/6 & 4.167$\pm 0.153$  \\
20.9 & 0.632$\pm 0.014$ & 0.510$\pm 0.038$ & 2.4/6 & 3.974$\pm 0.190$  \\
22.9 & 0.597$\pm 0.013$ & 0.372$\pm 0.037$ & 0.7/6 & 3.749$\pm 0.188$  \\
24.0 & 0.669$\pm 0.016$ & 0.478$\pm 0.043$ & 5.5/6 & 4.205$\pm 0.206$  \\
25.0 & 0.649$\pm 0.015$ & 0.517$\pm 0.039$ & 4.1/6 & 4.076$\pm 0.191$  \\
26.1 & 0.717$\pm 0.016$ & 0.583$\pm 0.042$ & 3.8/6 & 4.508$\pm 0.190$  \\
27.1 & 0.730$\pm 0.015$ & 0.503$\pm 0.042$ & 2.0/6 & 4.585$\pm 0.197$  \\
28.1 & 0.732$\pm 0.016$ & 0.576$\pm 0.044$ & 5.7/6 & 4.598$\pm 0.199$  \\
29.2 & 0.743$\pm 0.017$ & 0.557$\pm 0.047$ & 4.1/6 & 4.669$\pm 0.198$  \\
31.2 & 0.763$\pm 0.017$ & 0.683$\pm 0.047$ & 4.9/6 & 4.796$\pm 0.200$  \\
33.3 & 0.714$\pm 0.014$ & 0.652$\pm 0.037$ & 1.8/6 & 4.486$\pm 0.155$  \\
35.4 & 0.740$\pm 0.014$ & 0.624$\pm 0.037$ & 5.9/6 & 4.648$\pm 0.204$  \\
37.4 & 0.724$\pm 0.021$ & 0.690$\pm 0.056$ & 1.5/6 & 4.549$\pm 0.259$  \\
39.5 & 0.888$\pm 0.017$ & 0.834$\pm 0.046$ & 7.3/6 & 5.580$\pm 0.217$  \\
43.5 & 0.939$\pm 0.015$ & 0.857$\pm 0.040$ & 3.2/6 & 5.900$\pm 0.240$  \\
\end{tabular}
\end{table}

\newpage
\begin{table}

\caption{The $\chi^2/data$ for the present differential cross sections vs
pion kinetic energy for recent VPI SP94 and SM95
solutions\protect\cite{SAID}.  SP94 does not include the present data
while SM95 includes them.  Norm is a common normalization factor
determined from the SP94 and SM95 solutions.}
\label{tab3}

\begin{tabular}{ccccc}
  $T_{\pi}$   &\multicolumn{2}{c}{SP94}&\multicolumn{2}{c}{SM95} \\
 (MeV) & Norm &$\chi^2/data$& Norm &$\chi^2/data$ \\
\tableline
18.8 & 1.04 &  8/6 & 1.03 &  7/6 \\
20.9 & 0.98 &  3/6 & 0.98 &  3/6 \\
22.9 & 0.93 & 14/6 & 0.93 & 14/6 \\
24.0 & 1.01 & 11/6 & 1.01 & 11/6 \\
25.0 & 0.98 &  8/6 & 0.97 &  8/6 \\
26.1 & 1.04 &  7/6 & 1.04 &  7/6 \\
27.1 & 1.05 & 17/6 & 1.05 & 16/6 \\
28.1 & 1.04 & 12/6 & 1.03 & 11/6 \\
29.2 & 1.04 & 14/6 & 1.04 & 13/6 \\
31.2 & 1.04 &  7/6 & 1.04 &  7/6 \\
33.3 & 0.96 &  4/6 & 0.96 &  3/6 \\
35.4 & 0.98 & 13/6 & 0.98 & 12/6 \\
37.4 & 0.94 &  4/6 & 0.94 &  4/6 \\
39.5 & 1.08 & 13/6 & 1.07 & 12/6 \\
43.5 & 1.09 & 13/6 & 1.09 & 13/6 \\
\tableline
\multicolumn{2}{c}{All energies}&147/90&      &142/90 \\
\end{tabular}
\end{table}

\newpage
\begin{table}
\caption{Comparison of $\chi^2/data$ for the present and World 
total and differential cross sections for the recent VPI SP94 and SM95
solutions\protect\cite{SAID} and previous solutions BU93\protect\cite{bugg}
and HI84\protect\cite{hiro}.  World data is selected for the energy range
18 -- 44~MeV covered by the present experiment.}
\label{tab4}
\begin{tabular}{ccccc}
Solution &\multicolumn{2}{c}{$d\sigma/d\Omega_{c.m.}$}&
\multicolumn{2}{c}{$\sigma_{tot}$}\\
  & Present & World & Present & World\\
\tableline
SM95 & 142/90 & 297/98 & 12/15 & 38/21 \\
SP94 & 147/90 & 292/98 & 12/15 & 40/21 \\
BU93 & 176/90 & 319/98 & 52/15 & 45/21 \\
HI84 & 332/90 & 364/98 & 93/15 & 73/21 \\
\end{tabular}
\end{table}
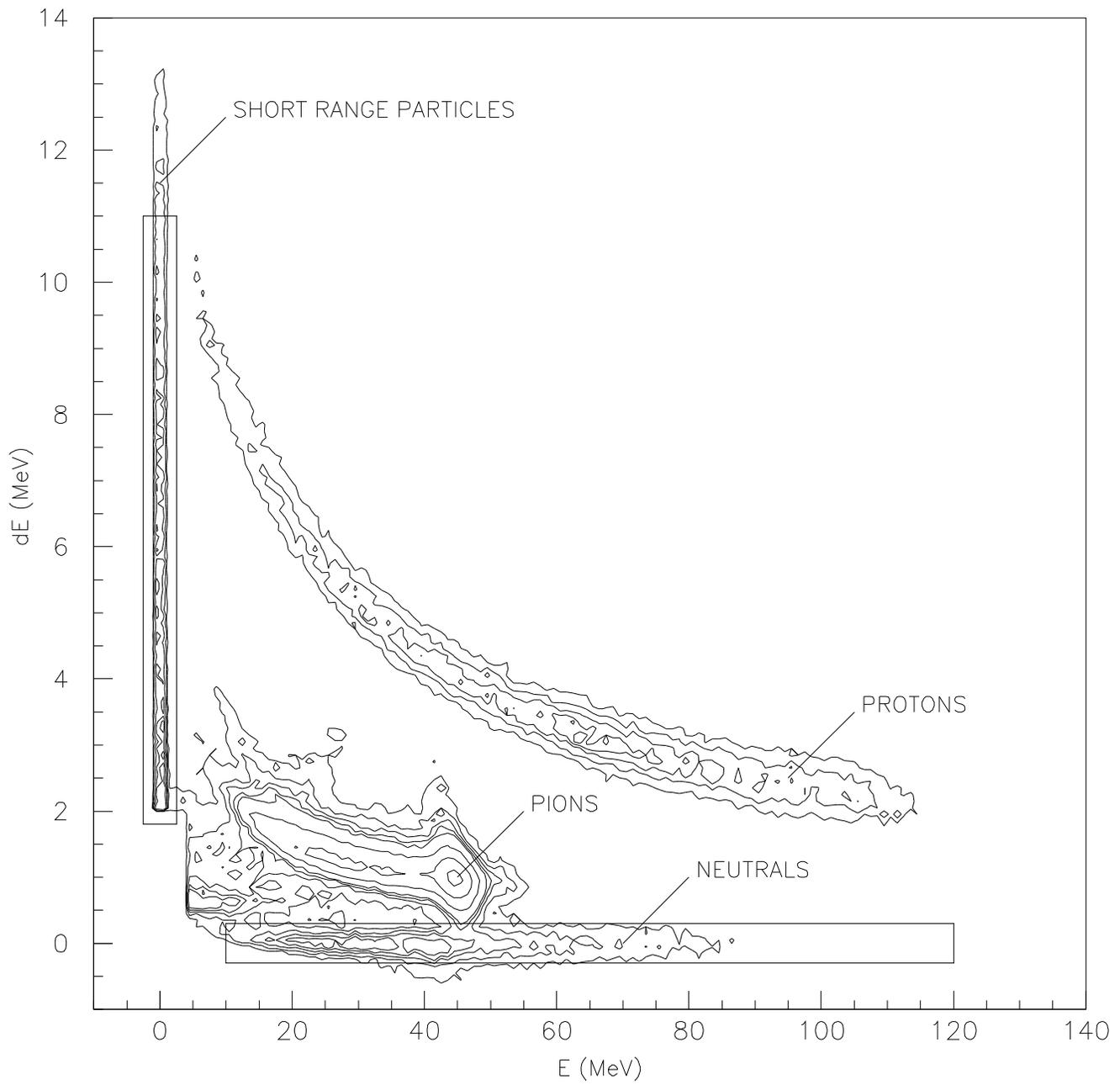
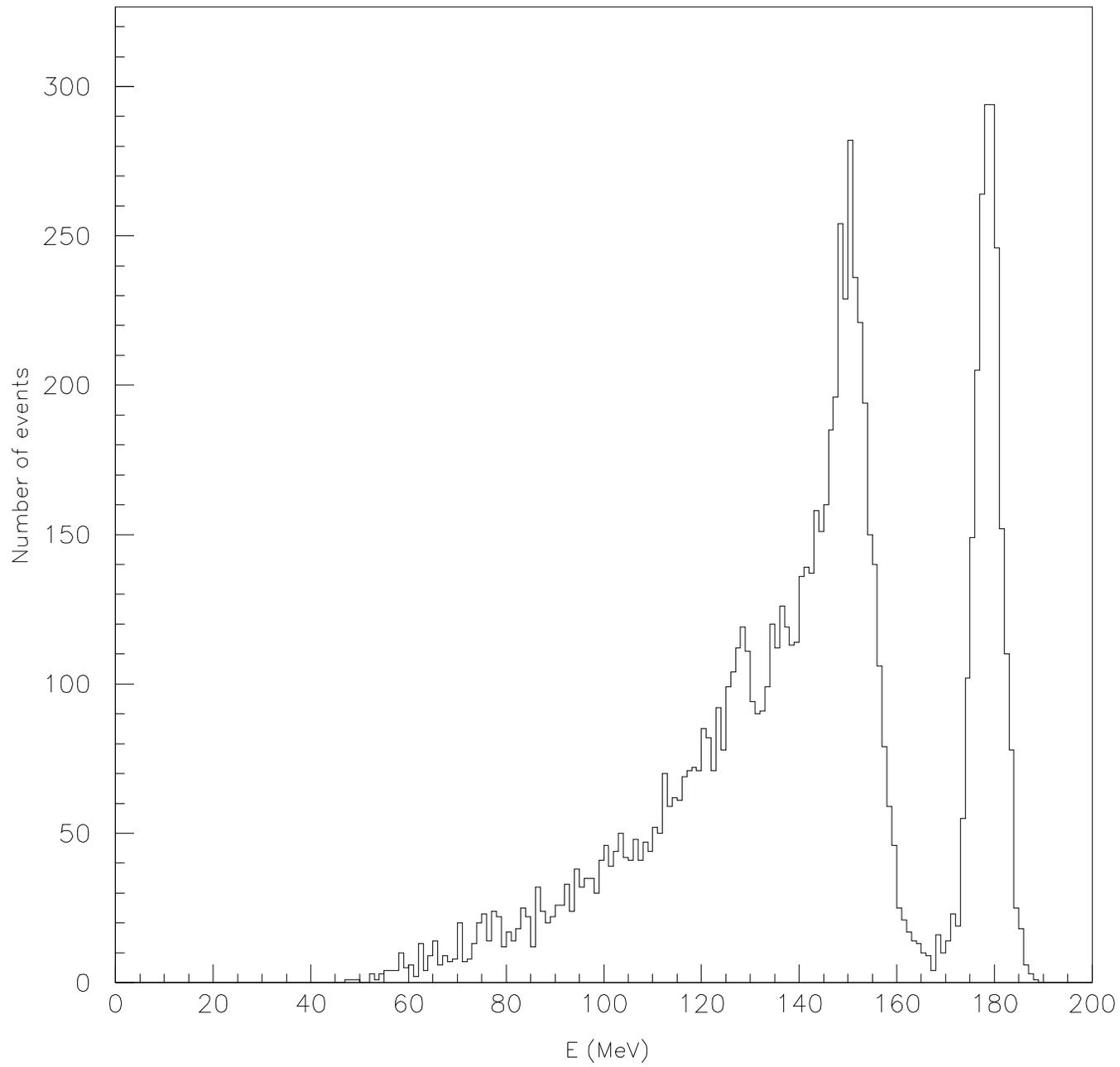
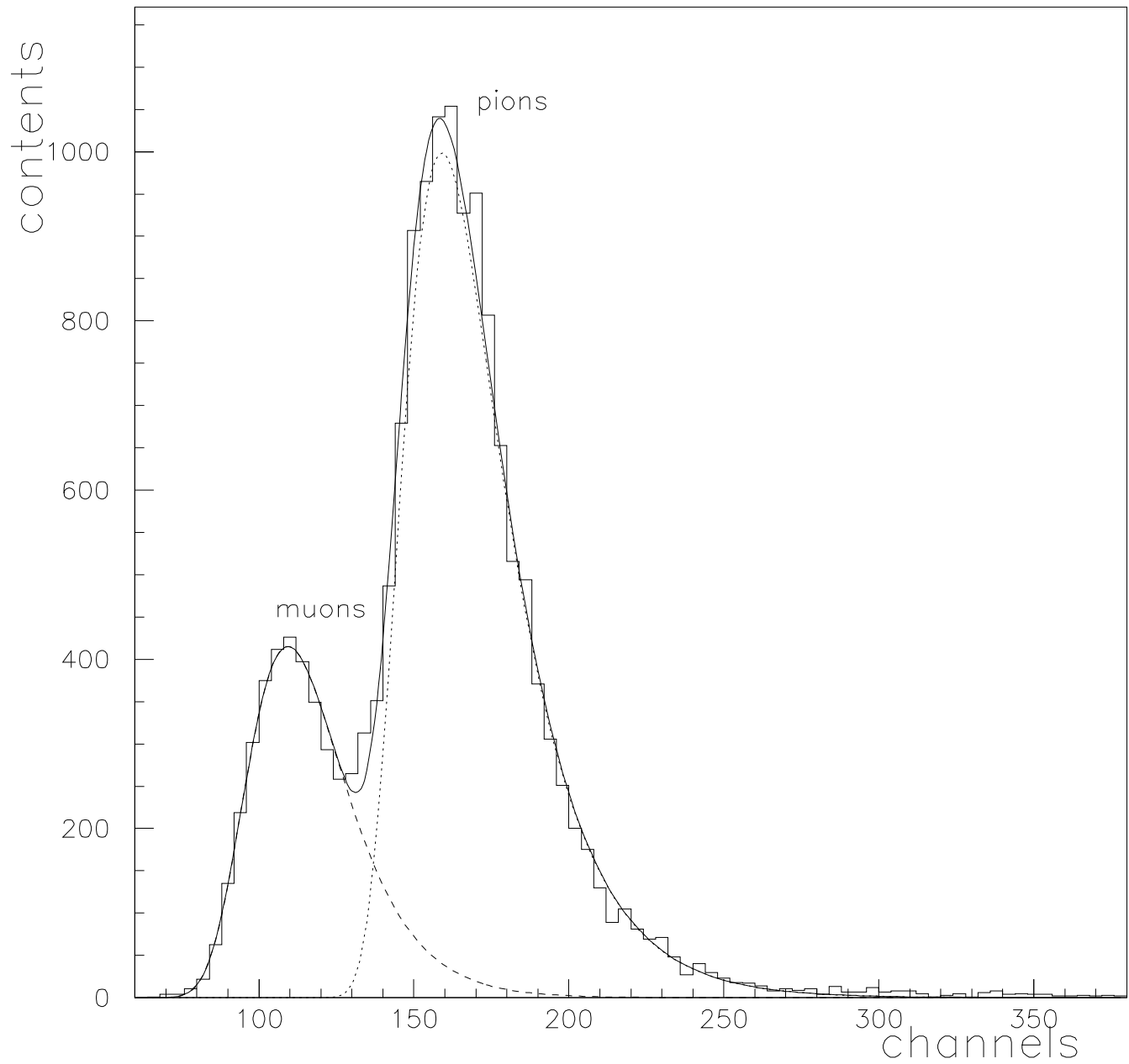
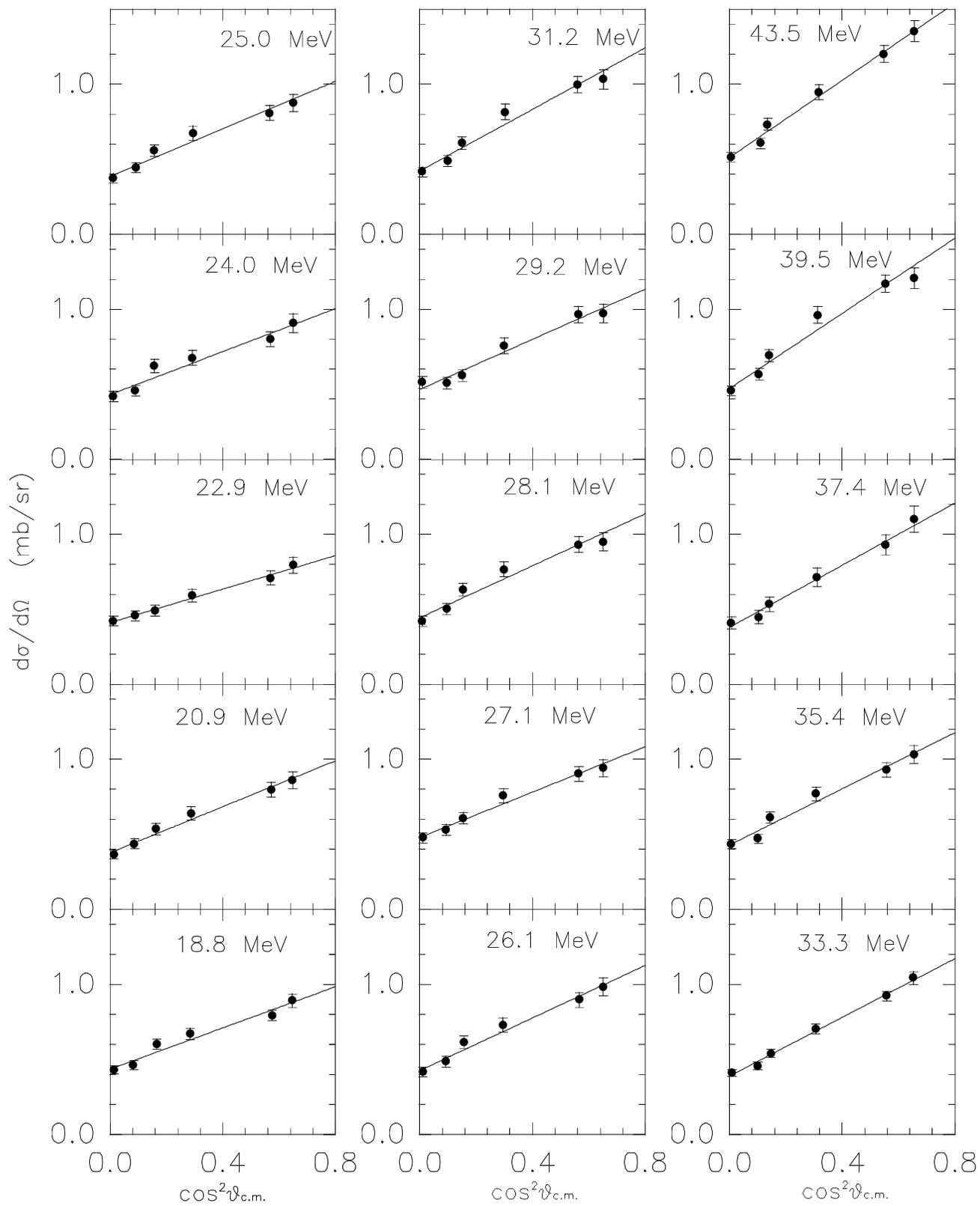
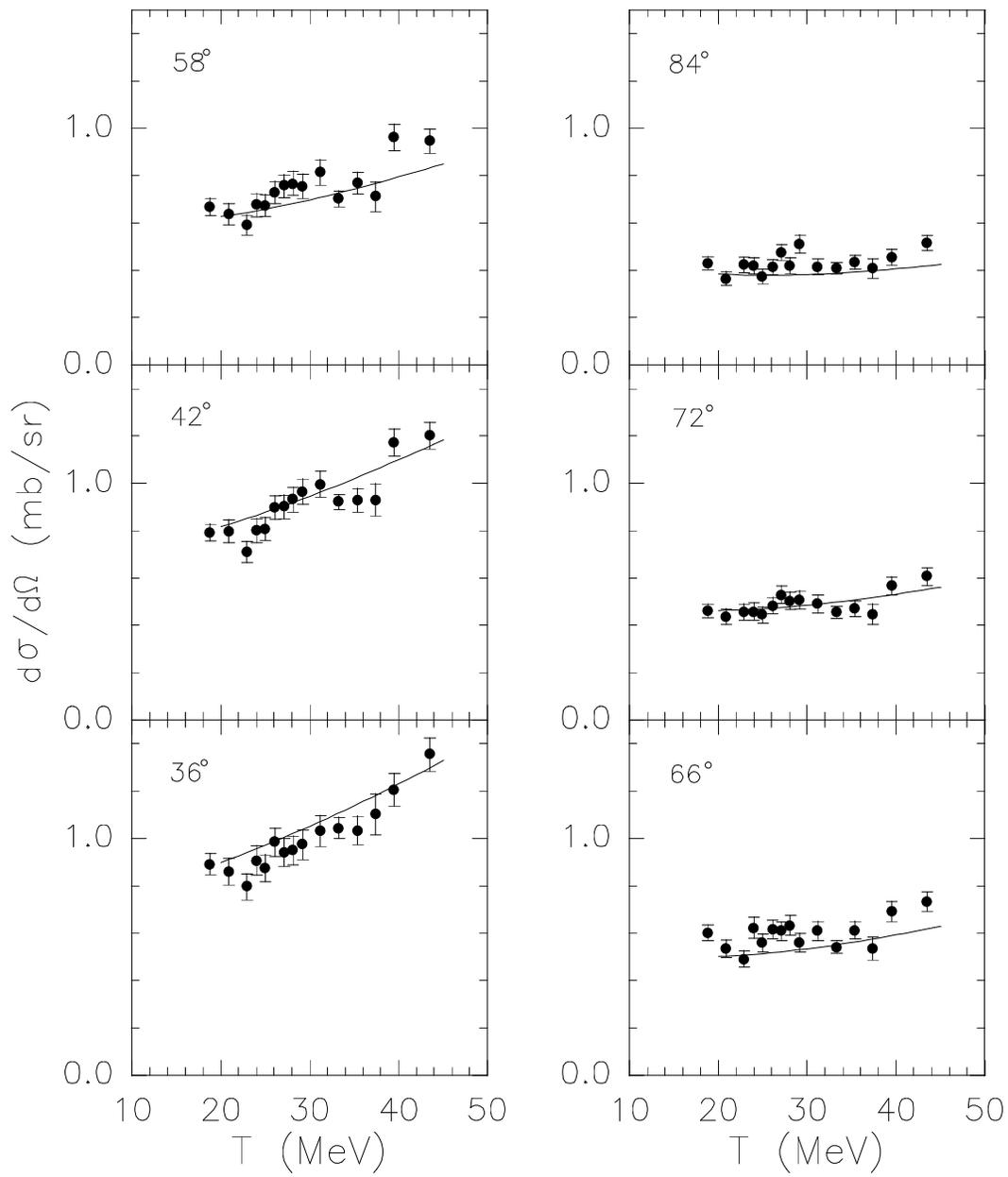
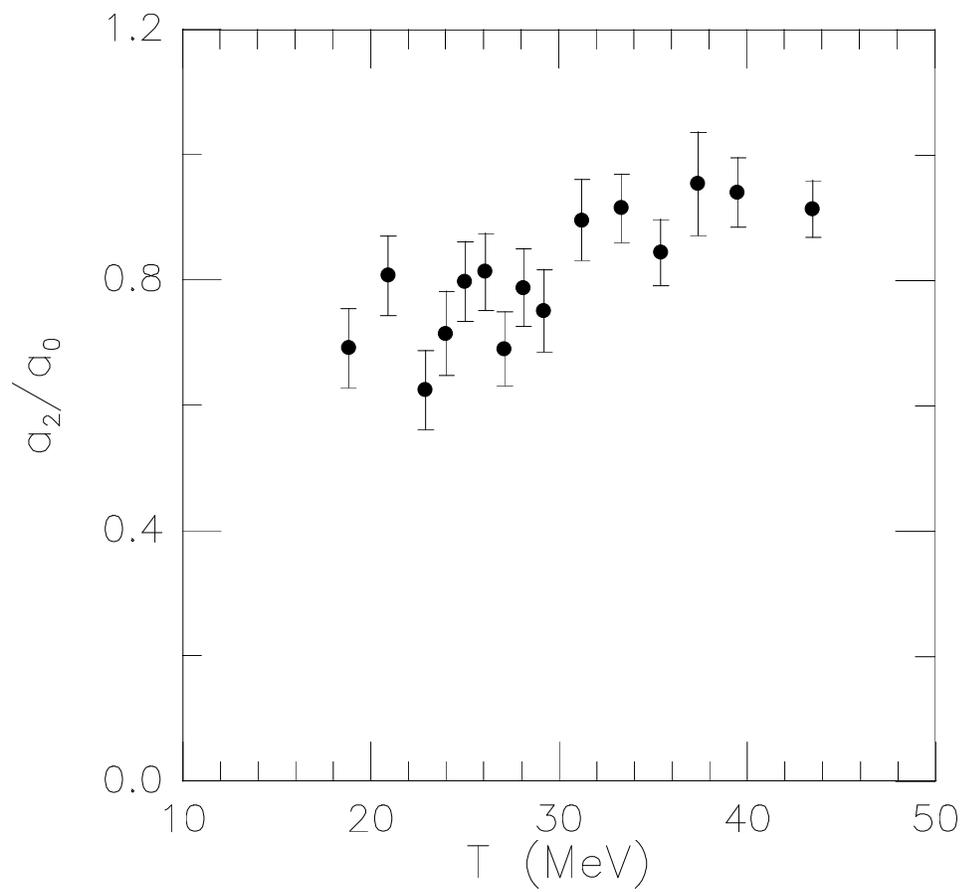
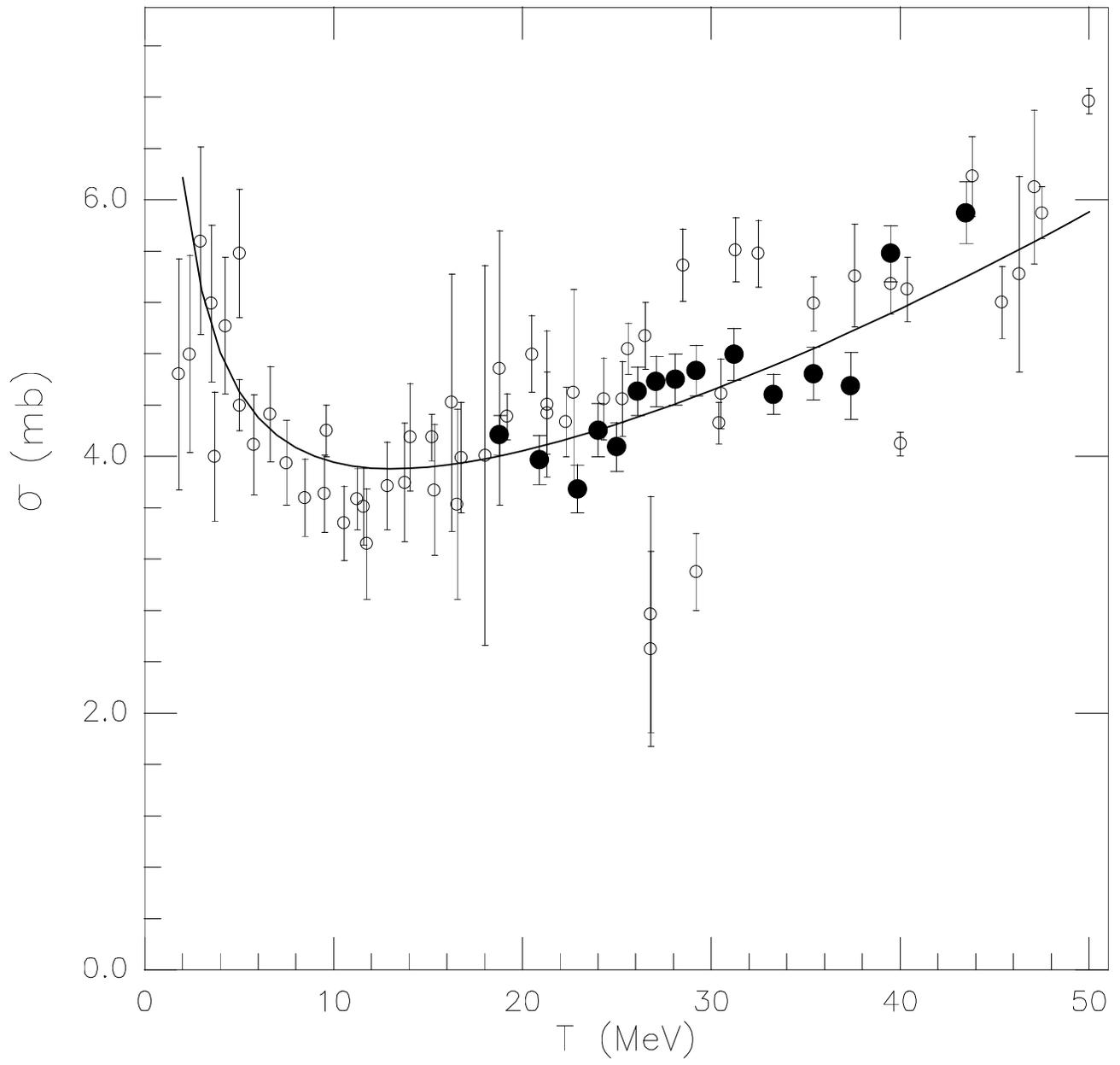
\end{document}